\documentclass[12pt]{elsarticle}
\usepackage{fullpage}
\usepackage{graphicx}
\usepackage{amsmath}
\usepackage{amssymb}
\usepackage{wrapfig}
\usepackage{amsfonts}
\usepackage{verbatim}
\usepackage[normalem]{ulem}
\usepackage[toc, page]{appendix}
\usepackage[switch, modulo]{lineno}
\usepackage[headsep=2cm]{geometry}
\usepackage{fancyhdr}
\fancypagestyle{empty}{
\fancyhf{}
\fancyhead[L]{\today}
\fancyhead[R]{\textbf{Version:} v16c}
}

\newcommand{\LAPPDTM}{LAPPD\textsuperscript{TM}~}
\graphicspath{{Figures/}}

\begin{document}

\title{ Low-Dose High-Resolution TOF-PET Using Ionization-activated
Multi-State Low-Z Detector Media}

\author[UC]{J. F. Shida}
\author[UC]{E. Spieglan}
\author[Comp]{B. W. Adams}
\author[UC]{E. Angelico}
\author[UC]{K. Domurat-Sousa}
\author[UC]{A. Elagin}
\author[UC]{H. J. Frisch}
\author[RAD]{P. La Riviere}
\author[PME]{A. H. Squires}
\address[UC]{Enrico Fermi Institute, The University of Chicago, 5640 S Ellis Ave, Chicago, IL 60637}
\address[Comp]{Quantum Optics Applied Research, Naperville, IL 60564}
\address[RAD]{Department of Radiology, The University of Chicago, Billings Hospital, P220, 5841 South Maryland Avenue, MC2026, Chicago, IL 60637}
\address[PME]{Pritzker School of Molecular Engineering, The University of Chicago, 5640 South Ellis Avenue, Chicago, IL 60637}

\begin{abstract}
 We propose PET scanners using low atomic number
media that undergo a persistent local change of state along the paths
of the Compton recoil electrons. Measurement of the individual
scattering locations and angles, deposited energies, and recoil
electron directions allows using the kinematical constraints of the
2-body Compton scattering process to perform a statistical
time-ordering of the scatterings, with a high probability of precisely
identifying where the gamma first interacted in the detector. In these
cases the Line-of-Response is measured with high resolution, determined
by the underlying physics processes and not the detector segmentation.
There are multiple such media that act through different mechanisms. As
an example in which the change of state is quantum-mechanical through a
change in molecular configuration, rather than thermodynamic, as in a
bubble chamber, we present simulations of a two-state photoswitchable
organic dye, a `Switchillator', that is activated to a
fluorescent-capable state by the ionization of the recoil electrons.
The activated state is persistent, and can be optically excited
multiple times to image individual activated molecules. Energy
resolution is provided by counting the activated molecules. Location
along the LOR is implemented by large-area time-of-flight MCP-PMT
photodetectors with single photon time resolution in the tens of ps and
sub-mm spatial resolution. Simulations indicate a large reduction of
dose.
\end{abstract}

\maketitle
\thispagestyle{empty} 



\section{Introduction}
\label{Introduction}
 As an outcome of developing large area
MCP-PMT-based pico-second time-of-flight
systems~\cite{history_paper,timing_paper,Limitations_Workshop_2011}
for identification of charged particles in high-energy
colliders~\cite{OTPC_paper,Philadelphia_Drifting_Light_proceedings} and
imaging of low-energy electrons in neutrino
physics~\cite{Andrey_paper_1,Andrey_paper_4,ANNIE}, we have explored
PET scanners based on large-area MCP-PMT photodetectors viewing low
atomic number scintillating media~\cite{PET_patent}. The goal is to
resolve the location, recoil direction, and energy of the chain of
successive Compton-scattered electrons in the detector medium.
Simulations show that reconstructing the recoil electron tracks enables
a time-ordering of the Compton scatterings in the detector, with a high
probability of precisely identifying the site of the first interaction
of each gamma in the detector. Connecting the locations of the first
interaction for both gammas provides a precise determination of the
line-of-response (LOR).

Positron-Emission Tomography (PET) uses the selective uptake of
biologically active molecules labelled with one or more radioactive
positron-emitting tracers to image sites of biological
activity~\cite{Vandenberghe_Moskal_Karp_review_2020,Vaquero_Kinehan_review_2015,Phelps_Cherry_Dahlbom_book_2006}.
The emitted positron multiple-scatters and then annihilates with an
electron, emitting approximately back-to-back gamma rays, each with the energy of the electron mass. 
Time-of-flight PET (TOF-PET) uses the
difference of arrival times of the two gamma rays to constrain the
annihilation site along the Line-of-Response (LOR), the line connecting
the measured gamma ray interaction points in the
detector~\cite{Phelps_Cherry_Dahlbom_book_2006}. Here we describe an
optical system based on large-area MCP-PMT-based photodetectors to
provide cm-scale TOF resolution along the LOR.

The field is undergoing rapid development. Highly sophisticated
whole-body scanners with large solid angle acceptance have recently
been
developed~\cite{Vandenberghe_Moskal_Karp_2020_Whole_Body_PET_2020,Cherry_Explorer_scattering_2019}.
TOF-PET with sub-nanosecond coincidence is being developed ~\cite{Lee_Levin_100ps_2021}, and a challenge is currently in place to develop sub-10 ps TOF resolution ~\cite{lecoq_2019}. A particularly attractive
method exploiting ultra-fast timing using MCP-PMT photodetectors and
Cherenkov light in
pre-radiators~\cite{Credo,Ohshima,Anatoly_TestBeam_2010} is being
developed by Cherry et al. for higher spatial resolutions and lower
doses~\cite{Cherry_Hamamatsu_2021}. 

We had previously proposed whole-body scanners based on large-area psec
MCP-PMT photodetectors viewing high-Z
scintillator~\cite{Heejong_NIM_2010,Heejong_NIM_2011} and low-Z liquid
scintillator~\cite{PET_patent}.  A pioneering proposal by Moskal et al. emphasized the lower cost of low-Z scintillator-based PET ~\cite{Moskal_Organic_Scint_2012}.

Commercial PET scanners currently employ crystals containing high
atomic number elements to localize the gamma ray
interaction~\cite{Renzu,Derenzo}. The resolutions in time and space are
typically limited by the physical dimensions of the high-Z crystals and
their associated electronics~\cite{Moses_Fundamental_Limits}.

Moving from high-Z-inorganic to low-Z-organic media may allow optical
interrogation of the interactions in the detector at significantly higher resolution in both space and time,
with the goal of approaching the fundamental limits set by the
underlying physical processes. These processes include positron
emission, positron-electron annihilation, Compton scattering of gamma
rays and electrons, ionization energy loss, the generation of
scintillation light, Cherenkov emission, and the diffusion of
molecules.

There are candidate low-Z media that act through different mechanisms,
including those with a change in molecular states, and others in which
the change is thermodynamic, as in a bubble
chamber\cite{glazer_bubble_chamber}. For a specific example, we discuss
the potential for development of appropriate two-state photoswitchable organic dyes,
`Switchillators', that are activated to a fluorescent-capable state by
the ionization of the recoil electrons, and which offer additional prospects for synthetic fine-tuning at the molecular level to optimize the efficiency and kinetics of photoswitching and the spectroscopic properties of the active state ~\cite{Iwai_fluor_development_2020,super_resolution_Nevskyi_2018,super_resolution_Mockl_2020}.

A high-resolution spatial image of the Compton interactions also
enables a more precise resolution in Time-of-Flight (TOF), as the
transit times of the scintillation photons are predicted by the precise
location of the initial gamma interaction point. The development of
large-area multi-channel psec optical
 detectors~\cite{history_paper,timing_paper,Limitations_Workshop_2011,Incom_production}
allows the recording of multiple photons in time and space with sub-mm
space resolution and sub-50 psec time resolution. The ability to
measure the transit time of photons as well as the arrival location on
the detector face allows 3D spatial reconstruction (the `Optical Time
Projection Chamber',
OTPC~\cite{OTPC_paper,Andrey_paper_1,Oberla_thesis}). In the case of
the identification of charged particles, exploiting Cherenkov light in
a radiator on the face of the MCP-PMT produces time resolutions below 5
psec~\cite{Credo,Ohshima,Anatoly_TestBeam_2010,Philadelphia_Drifting_Light_proceedings}.

The organization of the paper is as follows; Section~\ref{Low_Z}
describes analyzing the chain of successive Compton scatterings in a
low-Z scintillating medium in a generic TOF-PET detector to determine
the site of the first interaction of each gamma and the extraction of
the LOR at high resolution. Section~\ref{Switchillator} presents
results from simulations of an example implementation using two-state
photoswitchable organic dyes activated to a fluorescent-capable state
by the ionization of the recoil electrons. Section~\ref{Summary} summarizes the proposed system and suggests possibilities for future implementation and impact.
%
%
\section{Using Low-Z Materials to Measure the Line of Response}
\label{Low_Z}

\begin{figure}
  \centering
    \includegraphics[width = 0.4\textwidth]{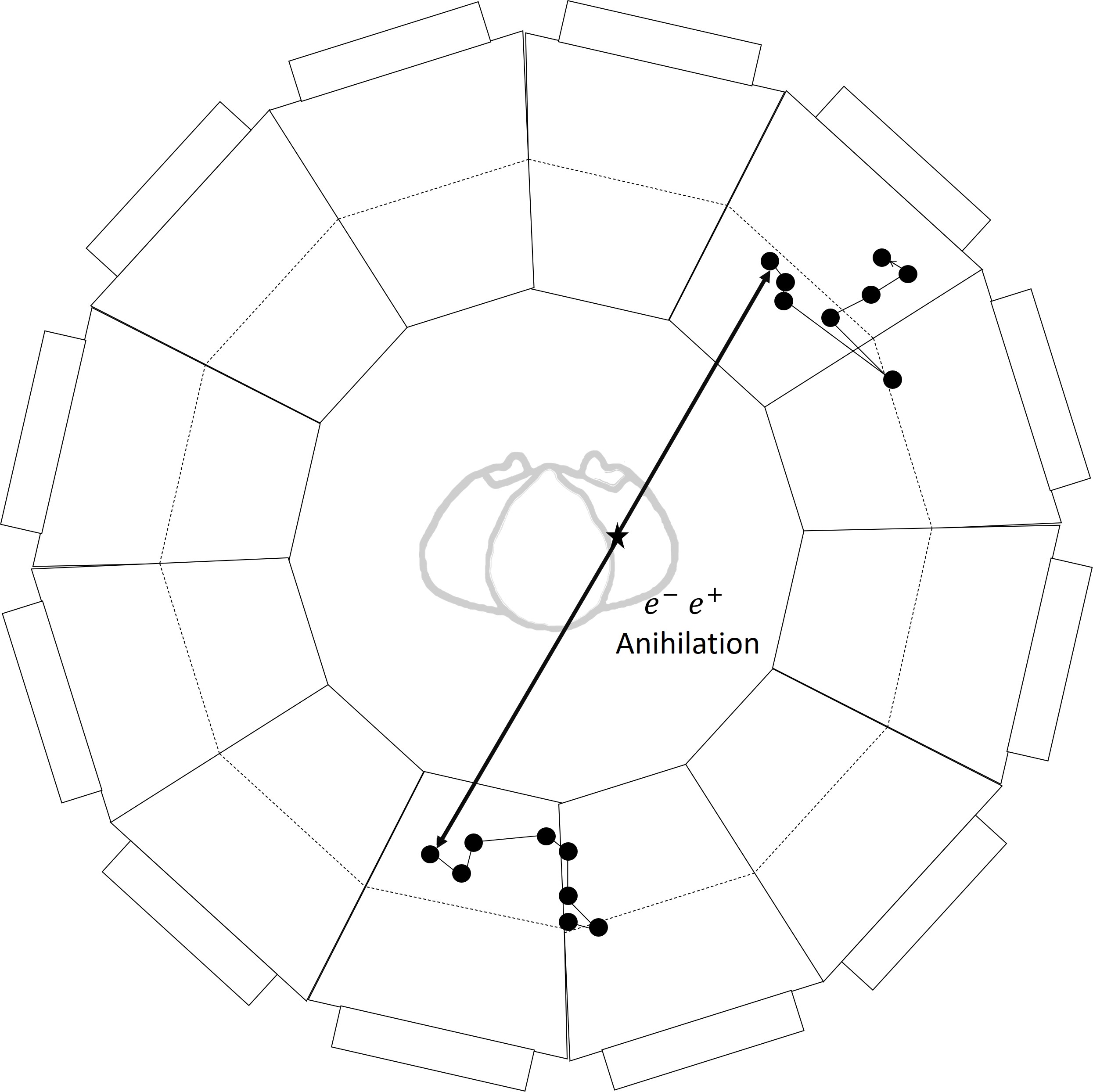}
    \caption{
   An axial view of an example modular TOF-PET whole-body scanner based on low-Z
   organic media. The patient is located on the axis of
    cylindrical rings of imaging modules covering the field-of-view.  Each of
    the two gamma-rays of a gamma-ray pair enters an imaging module and interacts via
    multiple Compton scatterings, producing recoil electrons that deposit energy
    along the recoil tracks.  An event simulated using the Kamland-Zen LAB-based
    scintillator~\cite{Kamland-Zen} as the medium is superposed to scale.}
  \label{fig:detector_end_view_Comptons}
\end{figure}

In low-Z gamma-ray detection the Compton cross-section dominates the
Photo-Electric (PE) effect, resulting in a chain of successive Compton
scatterings in the detector, each with monotonically decreasing energy of the scattering gamma.  In the majority of
events the energy-angle constraints of the 2-body Compton kinematics
can be used to identify the initial interaction site of the gamma in
the detector. The precision on the gamma trajectory is further improved
by using the ionization along the electron track to identify which end
of the track is the site of the Compton
scattering~\cite{Plimley_PhD_thesis}.

There are many detector media that change thermodynamic
phase~\cite{glazer_bubble_chamber} or quantum state or chemical state locally where
ionization energy is deposited. In the example detector discussed in
Section~\ref{Switchillator}, the recoil electron tracks are recorded by
switching between two states of the molecules of a photoswitchable
fluorescent dye, a
`Switchillator'~\cite{Uno_Irie_2011,Irie_2014,Irie_2017,Kashihara_2017}.
The locations and number of activated molecules in the chain of
successive Compton scatterings are optically recorded by repeated
exposures of the activated Switchillator to excitation light that create fluorescent emission until the Switchillator eventually reverts to its inactive molecular conformation.”

The identification of the first Compton electron track is based on: the
relative fluorescence (`brightness') of the depositions (`clusters') and the distances and angles between successive clusters. The kinematic constraints of
Compton-scattering~\cite{Compton} among clusters on the reconstructed
gamma ray trajectory are used to trace the trajectory of the gamma as it scatters. 
The starting point of the first-scattered
electron track corresponds to the interaction point of the primary
gamma from the annihilation, and hence one end of the LOR. The location
can be determined from the evolution of ionization and scattering
along the electron track~\cite{Plimley_PhD_thesis}.
Figure~\ref{fig:detector_end_view_Comptons} shows an end view of the
conceptual cylindrical whole-body PET scanner used in the simulation
studies.

\subsection{The Gamma Trajectory: Successive Compton Scatterings}
\label{Compton_interactions}

The annihilation of a positron with an atomic electron produces two
approximately back-to-back gamma rays, each with an energy of the
electron mass, 511 keV~\cite{acoplanar}. These interact with the active
detector medium via two processes, Compton scattering~\cite{Compton},
and the photoelectric effect. The two process have distinctly different
behavior with atomic number (Z) of the target material;  when a 511 keV
gamma ray first enters a typical low-Z organic
scintillator~\cite{Kamland-Zen} the ratio of the Compton to PE
cross-sections is on the order of $10^4$~\cite{cross_sections}.

We have used  Geant4~\cite{Geant4_2003,Geant4_2010} to simulate the
successive interactions of  511 keV gamma rays entering a volume
containing a typical low-Z organic liquid scintillator, in this case
the LAB-based scintillator developed by the Kamland-Zen
collaboration~\cite{Kamland-Zen}. The pattern of ionization produced by
the gamma is determined by the kinematics of successive
Compton-scatterings, with each scatter producing a gamma of lower
energy and a recoil electron in a two-body process fully constrained by
 conservation of energy and
momentum~\cite{Compton,Klein_Nishina,PDG_Groom_Klein_2019}. For each
interaction the location, energy, and trajectory of the (ionizing)
electron were recorded.  In the simulation no details of the detector
have been applied; a proposal for one such implementation that could in principle approach 100\% detection efficiency is described in
Section~\ref{Switchillator}.

\begin{figure}[ht]
  \centering
   \includegraphics[width=0.85\textwidth]{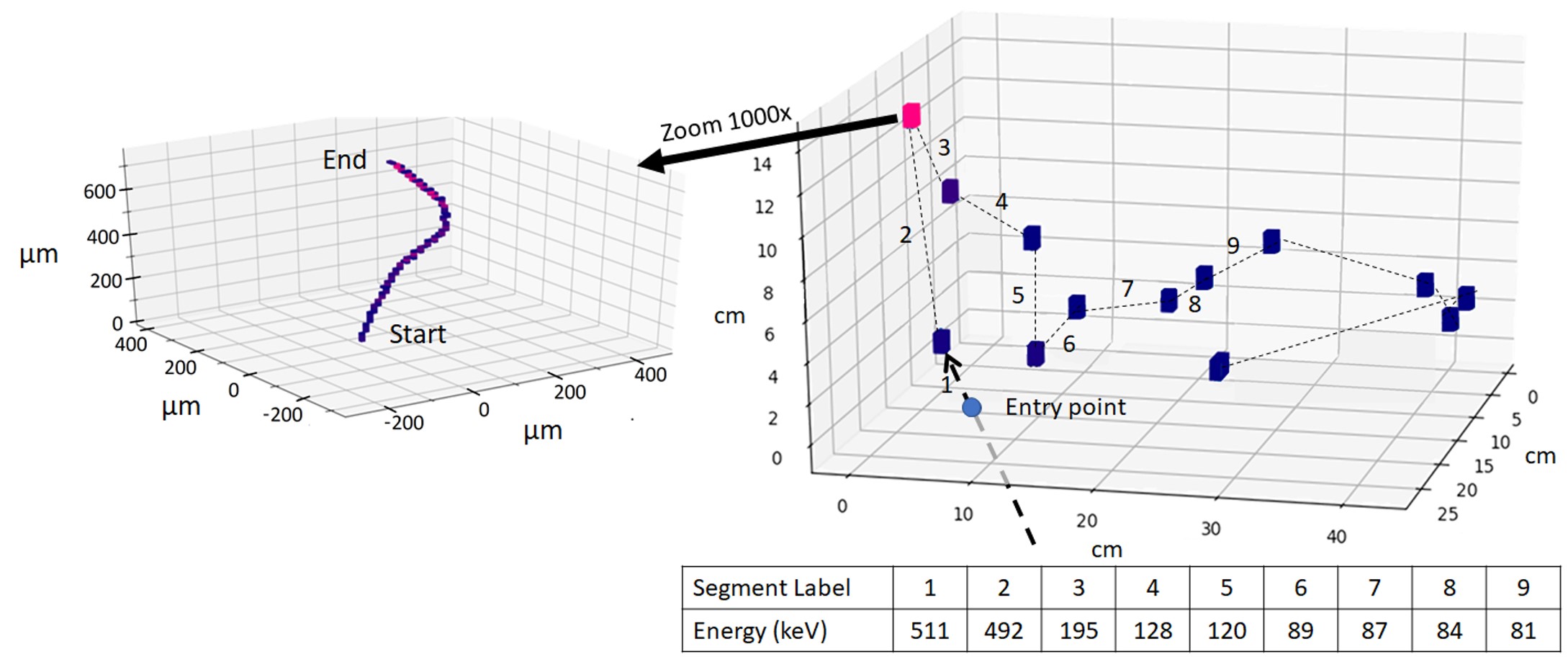}
  \caption{An image displayed in 1 cm pixels of the time-ordered energy depositions of a simulated 511
    keV gamma interacting in the Kamland-Zen scintillator~\cite{Kamland-Zen}.
    The gamma is shown entering
    the module from below.
    The inset on the left shows a map of the energy deposition from the most energetic
    Compton scattered electron in 10 micron voxels.
    The incident gamma energy for each of the enumerated Compton scatterings is
    tabulated below the image.}
  \label{fig:clusters}
\end{figure}

Figure~\ref{fig:clusters} shows the simulated trajectory of a 511 keV gamma ray
after entering a
 volume of LAB-based liquid scintillator~\cite{Kamland-Zen}. The right-hand panel corresponds to a
 high-resolution digital-camera image displayed in 1-cm pixels. The inset shows
    the energy depositions from the most energetic Compton scattered electron
    displayed in 10-micron voxels.

\begin{figure}[th]
  \centering
\includegraphics[width =1.00\textwidth]{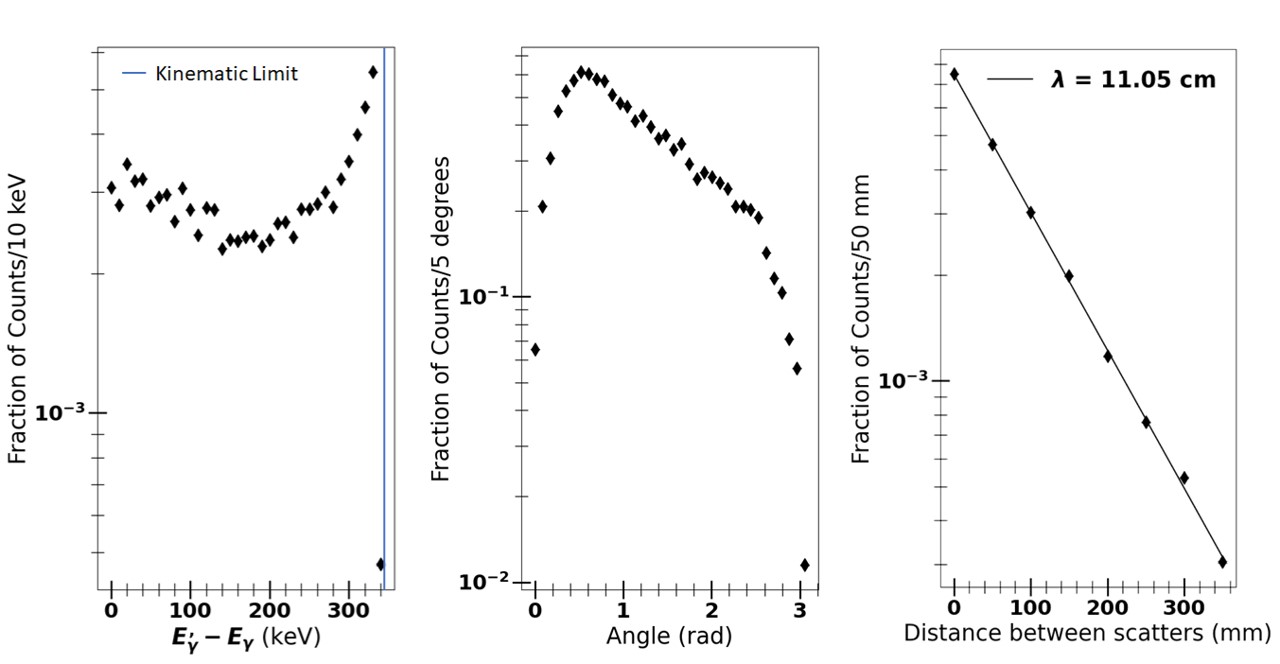}
\caption{A Geant4 simulation of the kinematics of the first scatter of
a 511 keV gamma ray in the Kamland-Zen low-Z scintillator. Left: the
distribution in the difference between the energies of the incident and
scattered gamma; Center: the distribution in scattering angle; Right:
the distribution in separation between the first and second
scatterings. The fitted line is the result of a fit of the separation
to the form $s = e^{-s/\lambda}$.}
  \label{fig:first_scatter_distributions}
\end{figure}

The distribution in separation between the first and second scatterings
is shown in the right-hand panel of
Figure~\ref{fig:first_scatter_distributions}. The separation is large
compared to the resolution of an optical system, allowing in most cases
the identification of individual clusters of energy corresponding to
the early scatterings in the chain. The distribution in the energy of
the scattering, i.e. the difference between the incoming and outgoing
gamma energies, is shown in the left-hand panel. The angle of
scattering in the first scatter is shown in the central panel.


If there is no in-patient scattering, the first Compton scatter of the
gamma ray in the detector is at the full energy of 511 keV. Successive
scatterings are at successively lower incident gamma energies. The
deposited energy from the first collision consequently has the highest
probability of being the largest (`brightest') in the chain.  The
simulation finds that largest ionization deposition occurs in the
first, second, and third scattering in 55\%, 25\%, and 11\% of gamma
interactions, respectively.

However, we can do better than 55\% accuracy in identifying the first interaction point if we use additional Compton kinematic constraints. The energy and direction of a scattered Compton electron are fully
constrained for a given incoming gamma ray energy and scattering angle.
One can consequently fit the observed array of locations and energies
for the time-ordering and the incoming gamma energy and direction. For
example, we find that a $\chi^2$ fit to the energies and angles of
the six permutations of the three brightest clusters finds the correct
time-ordering 86\% of the time~\footnote{ In the case of
misidentification we find an rms transverse resolution of 8 cm,
producing a low-contrast background to features with scales of the 40
micron transverse resolution. See Section~\ref{Imaging}.}.


The processes of energy loss by the scattered electron allow a
statistical determination of which end of the observed electron track
is the point of the $\gamma$e scattering, determining the recoil
direction~\cite{Plimley_PhD_thesis}. A constrained fit to the Compton
kinematics of the $\gamma$e scattering angles and $\gamma$ energy loss of
the chain of scatterings depends on this direction. We find that 96\%
of the electron tracks in largest energy deposition have the correct
end identified as the electron origin.

\subsection{Forming the Line of Response and Rejection of In-patient Scattering}
\label{Forming_LOR}

As shown in Figure~\ref{fig:detector_end_view_Comptons}, the two gamma
ray trajectories define the LOR, on which the annihilation point will
lie in the case of no intervening scattering in the patient or inert
detector material.  Using the Compton scattering equations to predict
which energy deposition was the first will correctly define the line of
response.  With as few as two detected Compton scatterings in the
detector, the simulation reconstructs the LOR correctly $\approx$60\%
of the time. The transverse resolution for each LOR is consequently on the order of the uncertainty of the transverse position of the scattering sites on each end.
Including more scatterings and more information ($\phi$-dependence,
distance between sites), and using more sophisticated techniques such
as machine learning~\cite{machine_learning_in_PET} will improve the accuracy.

The LOR represents the central axis of a prior probability
  density (dubbed a `needle' due to its elongated shape) for the true location of the $e^+e_-$ annihilation.
  For image reconstruction, this density distribution is modelled as Gaussian in both the transverse and
longitudinal directions.
  The transverse resolution, $\sigma_T$, is set by the spatial resolution on the end-points of the
  LOR of the optical system as described above, and for the Switchillator implementation described in
  Section~\ref{Switchillator} would be on the order of 40 microns. The longitudinal
  resolution, $\sigma_L$, is set by the resolution of the TOF system, and for the Switchillator
  implementation  would be on the order of 1.2 cm.

\begin{figure}[hbt]
  \centering
   \includegraphics[width=0.80\textwidth]{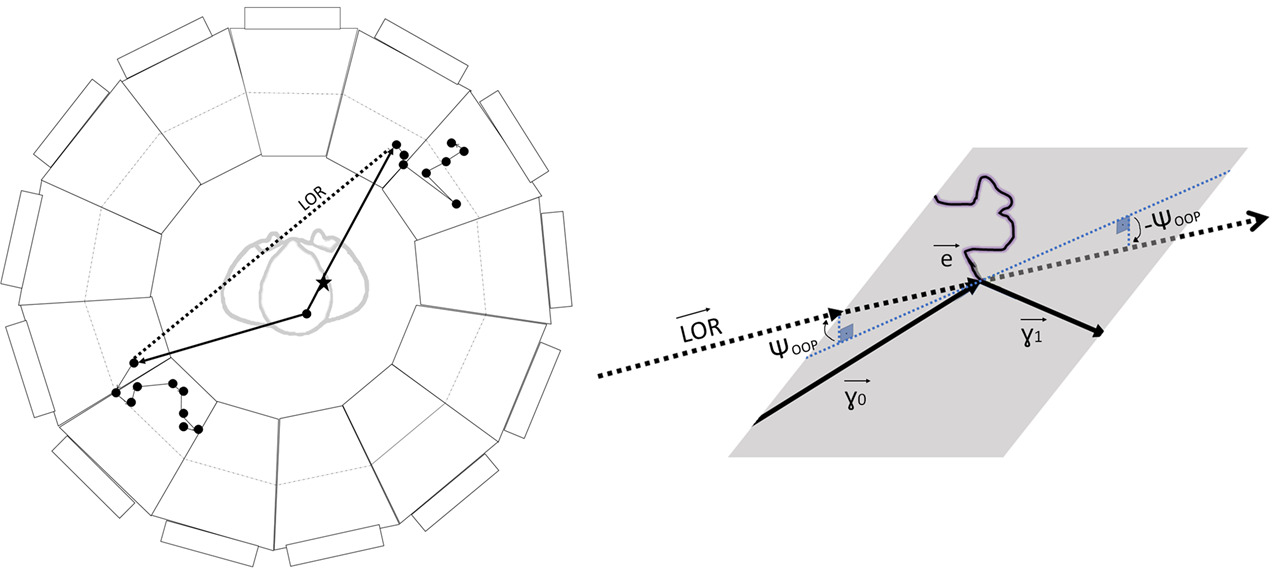}
   \caption{The LOR in the case of in-patient gamma-ray scattering.
 The positron annihilation point is marked by a star. The true gamma trajectories are represented by solid lines, and the
 reconstructed LOR connecting the two measured gamma interaction points by a dashed line.
 The inset shows that a gamma-ray scattering in
 the patient produces a non-zero
 component of the LOR out of the plane defined in the detector module
 by the scattered electron and gamma, i.e.
 $(\vec{e}\times\vec{\gamma_1})\cdot\vec{LOR}\ne 0$.}
  \label{fig:forming_LOR_scattering}
\end{figure}

A precision measurement of the incoming gamma energy can be used to
discriminate against scattering in the patient and inert material
before the active
volume~\cite{Vandenberghe_Moskal_Karp_review_2020,Phelps_Cherry_Dahlbom_book_2006,
inpatient_scattering_Hosokawa_2017}.
In addition, since the Compton process is 2-body, there should be no
measurable component of the LOR out of the scattering plane defined by
the recoil electron and the scattered gamma\footnote{Furthermore, the two gammas are quantum entangled and predominantly arisefrom  the  singlet  state  of  positronium.  The  Compton  scattering  cross-sectiondepends on polarization with a cos $\phi$ dependence. A full statistical analysis ofthe  cluster  ordering  can  include  this  as  a  prior,  with  possible  estimation  andrejection of scattering background.}, as shown in
Figure~\ref{fig:forming_LOR_scattering}.

The positron is emitted with a beta-decay spectrum, which for $^{18}$F has
an endpoint at 634 keV and a mean of 250
keV~\cite{Conti_tracers_beta_spectrum_F18}. Multiple scattering and
ionization loss sculpt the positron range distribution into a
`cusp'-like distribution, with a typical FWHM for $^{18}$F of $\approx$50
microns~\cite{Positron_multiple_scattering_Blanco,Positron_multiple_scattering_Levin,Positron_multiple_scattering_Derenzo}.
We have included the effect of positron range by convolving the
annihilation point with the distribution of Fig. 3 of Blanco et
al.~\cite{Positron_multiple_scattering_Blanco}.


The initial positron momentum from nuclear beta-decay is not negligible
compared to the 511 keV energy of the gamma rays, and as a consequence
the motion of the center-of-mass of the electron-positron system
distorts the back-to-back distribution in the frame of the
annihilation~\cite{acoplanar,Angular_distribution_Harpen,Angular_distribution_Moskal}.
Two effects mitigate the effect on the resolution of the annihilation
point: 1) the positron typically annihilates at a lower energy due to
ionization, and 2) the two-gamma
emission axis and the positron momentum direction are uncorrelated. In
the cases where the two directions are approximately collinear and the
momentum transfer is small the transverse momentum component to the LOR
will be
small~\cite{Herraiz_Deep-Learning_Range_Correction_2020,Positron_range_Carter_2019}.
A definitive treatment of the effect remains to be
done~\cite{Joaquin_private_comm}.

\subsection{Implementing a Stable Sub-nsec TOF Coincidence Window}
\label{TOF_Coincidence}

The TOF system provides the event trigger, which initiates scanning and data acquisition for the switchable recording medium. The function of the real-time
coincidence window is to reduce the event rate so that data acquisition
is deadtimeless~\cite{Amidei_CDF_trigger_1988}. The real-time trigger
window can be refined off-line, and hence is not critical, being chosen
for high efficiency, stability, and ease of implementation in a
physically-extended many-channel system in which a stable tight window
(e.g. below 1 nsec) would be difficult to maintain.

To tighten the coincidence window well below 1 nsec requires a
system-wide precise time reference such as the IEEE1558 White Rabbit
standard~\cite{White_Rabbit}, which we have measured to be precise at
the 5 psec level over 10's of meters in a commercial off-the-shelf
implementation~\cite{Seven_Solutions}.

 Higher level triggers, operating off-line on the fully-reconstructed data after
calibrations have been applied, can then tighten the coincidence
window. For the coincidence, the times and positions of scintillation
photons at the MCP-PMT face are digitized and recorded. For example,
the PSEC4 waveform sampling system samples at 10-15 GS/s at a precision
of 10.5
bits~\cite{PSEC4,Oberla_Clermont_2014,timing_paper,PSEC4A}.
The optical system records multiple scintillation photons with a
resolution on each below 40 psec~\cite{timing_paper}. A fit to the
leading edge of the recorded waveform determines the time, dominated by
the signal-to-noise ratio and the number of samples on the leading
edge~\cite{Limitations_Workshop_2011}.

%
%
\section{A Concept for a Whole-Body TOF-PET Detector Based on Photoswitchable Organic Dyes}
\label{Switchillator}

\begin{figure}[th]
  \centering
  \includegraphics[width= 0.65\textwidth]{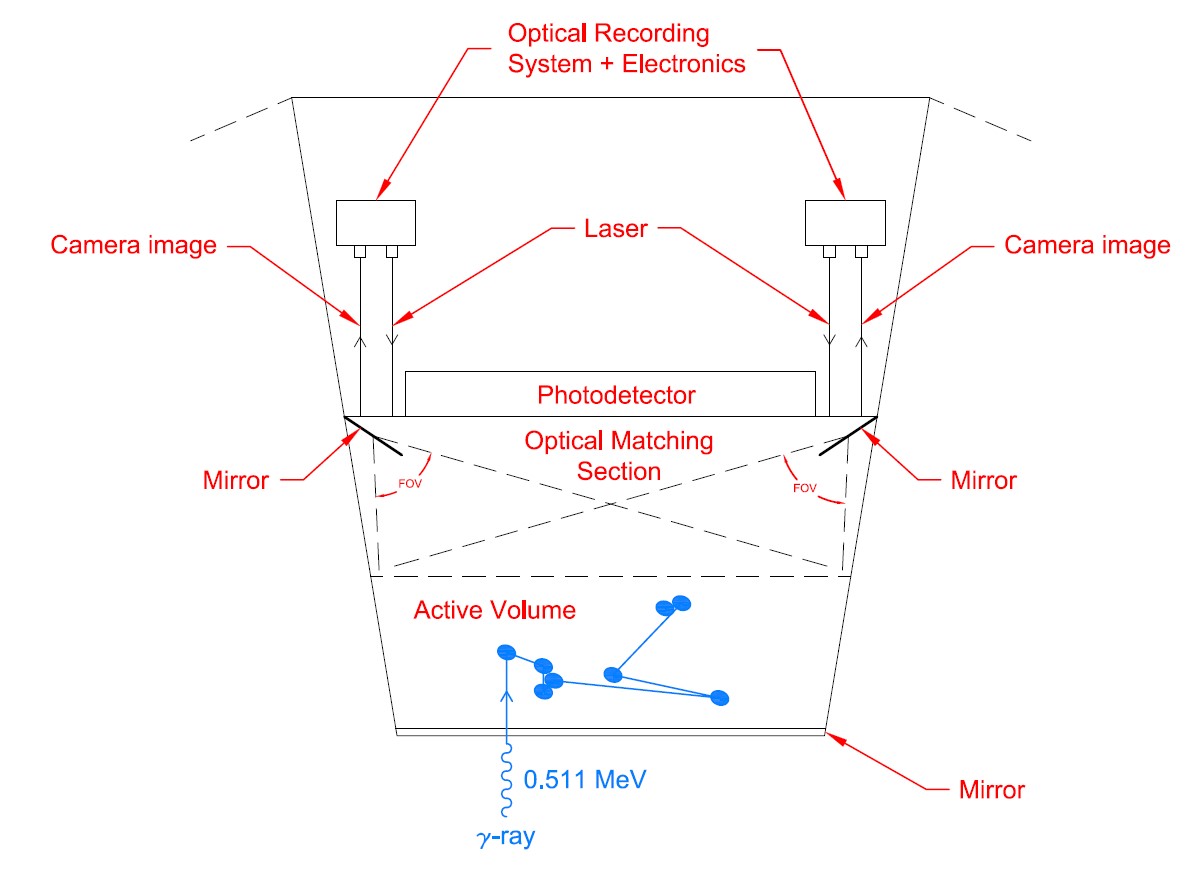}
  \caption{ An example module of a TOF-PET whole-body scanner based on Switchillator. One
  gamma of the pair from the annihilation enters through a front-side mirror. A large-area fast timing
  photodetector such as an \LAPPDTM is optically coupled through the back surface of the
  module. An optics system controls the scanning and focus of twin lasers to repeatedly
  excite Switchillator fluorophore molecules activated by ionization from Compton electron tracks
  in the Switchillator volume. Similarly, the optics system allows CCD cameras, which record
  after each laser pulse, to have a wide-angle stereo view  of the active Switchillator
  volume through the backside surface. A passive optical transition volume
  optically matched to the active volume allows the laser and camera
  systems to reach any point in the active volume.}
  \label{fig:Zack_Module}
\end{figure}

One possible phase-switchable recording medium that could be readily incorporated into a low-Z detection medium is a photoswitchable dye. Some of the energy deposited by each ionization event would be transferred by the medium to dyes in their inactive state, providing the necessary energy to chemically switch these dyes into a fluorescent, or active, molecular conformation. Combined with simple fluorescence excitation and detection optics, a Switchillator-based detector for low-Z TOF-PET could be implemented as proposed in Figure~\ref{fig:Zack_Module}. Each detection module includes an active volume containing organic
photoswitching fluorophores, a fast scintillator, and possibly other active
elements such as a triplet sensitizer.  Gamma rays from annihilations
in the sample enter the active volume through the module surface facing
the sample.

An optical subsystem to excite the activated Switchillator  molecules
is mounted on the back surface of the module. In the example shown in
Fig.~\ref{fig:Zack_Module} the excitation system is implemented using
twin diode lasers steered and focused by an optical system of
controllable mirrors and lenses to provide wide-angle stereo coverage
of the entire Switchillator volume. A similar optical subsystem using
twin CCD cameras images the fluorescence light from each excitation of
the Switchillator molecules by the laser system. To facilitate full
coverage with wide-angle stereo, each module has an optical transition
region transparent liquid with a matching index of refraction between
the active Switchillator volume and the module back face that supports
the optical systems.

TOF measurements using the initial scintillation light are provided by
a large-area MCP-PMT such as an \LAPPDTM mounted on the back module
surface.  In the simulation the internal surface of the module front
face has been made reflective to roughly locate energy depositions  and
to improve the time resolution by measuring the difference in
scintillation light transit times to the front and back module
surfaces. Electronics subsystems to support control, data acquisition,
calibration, and local data analysis, including a complete imaging of
the gamma ray trajectory in the module, are mounted on the back module
surface .

\subsection{Scintillator and Switchillator Properties}
The medium in the active volume consists of a fast scintillator, a
Switchillator fluorophore, and possibly other components to facilitate energy
transfer to the Switchillator molecules or to enable deactivation after
many excitation cycles. The required characteristics, and an estimate
of typical values, for the scintillator and Switchillator are given in
Table~\ref{tab:specs}.

\begin{table}
\label{tab:specs}

\centering
\begin{tabular}{|p{0.1in}|p{1.8in}|p{0.5in}|p{1.1in}|p{2.6in}|}
\hline\hline
 & Parameter & Symbol  & Value & Comment \\
\hline
\multicolumn{5}{|c|} {\bf Scintillator Properties} \\
\cline{1-5}
1 & Scintillation Yield & Y$_{scint}$ & $>2\times 10^3$ & \# of scintillation photons per MeV\\
2 & Scintillation Rise Time & $\tau_r$ & TBD & 1/e rise time of scintillation light\\
3 & Scintillation Decay Time & $\tau_d$ & TBD & 1/e decay time of scintillation light\\
\hline
\multicolumn{5}{|c|} {\bf Switchillator Properties} \\
\cline{1-5}
1 & Activation Yield & $Y_{act}$ & $>5\times 10^{3}$ & \# of ON fluorophores per MeV deposited\\
2 & Activation Wavelength & $\lambda_{act}$ & $< 400$ nm & Peak inactive to active wavelength \\
3 & Excitation Wavelength & $\lambda_{ex}$ & 350-650 nm & At max separation \\
4 & Dye Ratio & $Z_{dye}$ & $<10^{-12}$ & Ratio of rates of background activation to fluorescence at $\lambda_{ex}$ \\
5 & On-State Lifetime & $\tau_{ON} $ & $3\!\times\! 10^{-7}\!-\!10^{-1}$ s & 1/e Lifetime of ON fluorophores \\
6 & Fluorescence brightness & $\varepsilon \cdot \Phi_{fl}$ &$>\!10^{3}$/(M cm)& Rate of emission from active dye\\
7 & Mean Absorption Length& $\chi(\lambda_{ex})$ & $>6$ m & 1/e absorption length at $\lambda_{ex}$\\
8 & Emission Wavelength &  $\lambda_{fl}$ & 400-700 nm &  Wavelength of fluorescence light \\
9 & \# of photons per activated fluorophore & ${N_{fl}}$ & $>500$ & Mean \# of fluorescent photons extracted per fluorophore before deactivation\\

\hline\hline
\end{tabular}
\renewcommand\baselinestretch{1.0}
\caption{Physical properties and estimated associated desired values
for a diarylethene-based scintillator-switchillator detector medium
containing both a fast scintillator and the photoswitchable
Switchillator
molecules~\cite{Uno_Irie_2011,Irie_2014,Irie_2017,Kashihara_2017}.
Properties of particular concern and that are targeted for development
include the Dye Ratio (selectivity), On-state lifetime (resetting), and
Absorption Length (optical depth)~\cite{Eric_CPAD}. }
\end{table}

\subsection{Energy Resolution: Counting Scintillator Molecules as a
             Proxy for Deposited Ionization Energy}
\label{Energy_resolution} The Switchillator concept~\cite{Eric_CPAD} incorporates ionization-activated photoswitchable organic fluorophores ~\cite{Uno_Irie_2011,Irie_2014,Irie_2017,Kashihara_2017,Arai_Irie_2017,Barrez_2018} directly into the low-Z liquid scintillation material. 
When an ionizing particle, in this case a Compton electron scattered from a 511 KeV
gamma ray, deposits ionizing energy in the solvent of a liquid
scintillation material, excited states are produced in the solvent
molecules. These excited states in turn transfer their excitations to
the fluorophore solute molecules, converting them from an inactive state that
neither strongly absorbs visible light nor fluoresces, to an active state
that absorbs visible light and fluoresces with high efficiency. Once activated, the fluorophores can be excited by a light source and the resulting fluorescence can be
recorded to produce an image. The activated fluorophores can be repeatedly
excited by the excitation light source before either reverting to their
inactive state by a mechanism inherent in the material or excitation
process, or being externally reset.

    The ionizing tracks produced by electrons scattered along the trajectory of the gamma ray
    are recorded by the location and number of the activated fluorophore molecules. In typical
    solvents these molecules remain within about 10 microns of the path of the energy transfer for
    many milliseconds, enabling high resolution imaging of the successive individual gamma
    ray interactions along its trajectory in the scintillation material. The number of
    activated molecules is proportional to the deposited ionization energy, with a conversion
    factor that can be more than $10^4$ molecules per MeV~\cite{Eric_CPAD}.

By way of illustration, in toluene there are initially $1.35\times
10^4$ singlet excitations per MeV from 10 MeV
electrons~\cite{Baxendale_singlet_excitations_1973}. To achieve a 1\%
energy resolution at 511 KeV requires 20,000 activated Switchillator
fluorophore molecules per MeV. A process using only singlet excitations of
the solvent with an efficiency of 0.4 would provide an energy
resolution of $\approx 1.5$\%. There are $2.8 \times 10^4$ triplet
excitations per MeV; a process able to use both triplet and singlet
excitations at an efficiency of 0.4 would achieve an energy resolution
of 0.8\% at 511 KeV.

\subsection{Cluster Finding}
\label{Cluster_finding}

\begin{figure}[ht]
  \centering
 \includegraphics[width=0.80\textwidth]{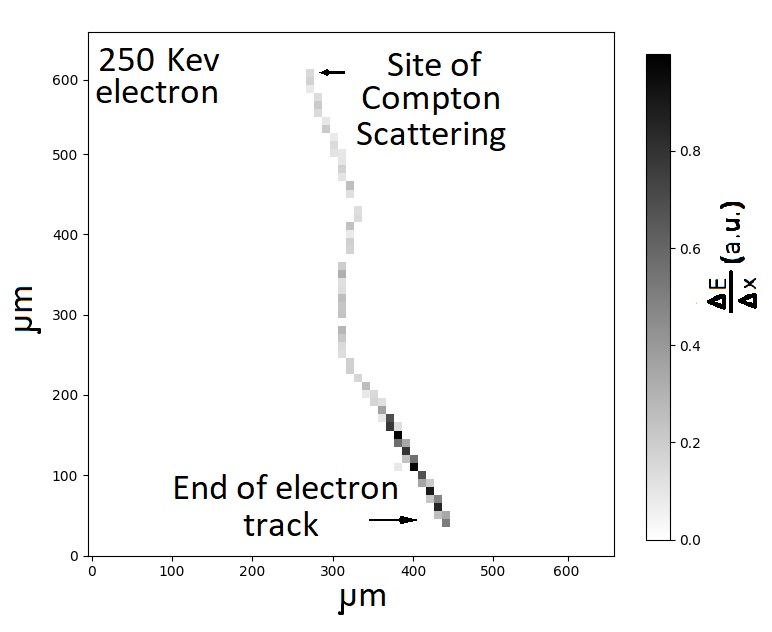}
  \caption{The output of a seed-shoulder cluster
algorithm~\cite{Amidei_CDF_trigger_1988} applied to a GEANT4 simulation
of a 250 keV electron imaged in pixels of 10 microns. The electron
deposits energy in the medium along its path. The most energy per voxel
is deposited as the electron ranges out. This can be used to identify
the position of the Compton scattering and the initial direction of the
electron at the start of the track.}
  \label{fig:Clustering}
\end{figure}

 Each interaction will produce a pattern of pixels containing activated molecules at
 a density above background. We have used a simple seed-shoulder
 algorithm~\cite{Amidei_CDF_trigger_1988} to form `clusters' of contiguous
 pixels based on deposited ionization energy. Figure~\ref{fig:Clustering} displays
 the output of the algorithm for the brightest cluster in
 Figure~\ref{fig:clusters}.

\subsection{Time-of-Flight System and Resolution}
\label{Switchillator_TOF}


The concept module shown in Figure~\ref{fig:Zack_Module} uses a method
of event reconstruction from the precise measurements of
time-of-arrival and position of Cherenkov light~\cite{OTPC_paper,
Oberla_thesis}. In analogy to the TPC technique of using drift
trajectories of electrons in a uniform electric field to make
time-sliced 3-dimensional images of ionization~\cite{Nygren_TPC}, we
have dubbed imaging by time-slicing the drifting of photons as the
`Optical Time Projection Chamber. In the conceptual design of
Figure~\ref{fig:Zack_Module}, each module supports a TOF system
consisting of a large-area MCP-PMT such as an \LAPPDTM and a mirror on
the inside of the entrance window. In the simulation results quoted
below the module sides were also reflective.


The design of the multi-level coincidence system of
Section~\ref{TOF_Coincidence} removes the need to narrow the real-time
coincidence window below that needed for deadtimeless operation of the
DAQ system~\cite{Amidei_CDF_trigger_1988}.

Off-line fits to the multi-photon photodetector pad or strip waveforms
are expected to provide time resolutions on the order of 40 psec or
below~\cite{Evan_thesis, timing_paper}, dependent on the speed and
light output of the scintillator and the details of the optical/MCP
system.

A simulation of the geometry of Figure~\ref{fig:Zack_Module} and an
infinitely fast scintillator with the light output of the Kamland-Zen
scintillator returns a 40 psec resolution, corresponding to a
longitudinal resolution $\sigma_L = 1.2$ cm (rms). With the  Kamland-Zen
scintillator the simulation returns a resolution of 40 psec.
Consequently a short coincidence window of a nsec or less, with the
width set by the difference in travel times needed to cover the
region-of-interest, can be used in the real-time trigger.

The TOF measurements can be further sharpened post-data acquisition by
using the locations and time-ordering of the clusters. However, while
promising, the actual performance will depend on the speed,
light-yield, and spectrum of the scintillator, the efficiencies and
performance of the light collection and detection system, and the
faithfulness of the reconstruction/cluster-ordering algorithms. In the
simulation we consequently use a conservative estimate of 40 psec
(sigma) for the TOF resolution.

\subsection{Constructing an Image from the Lines of Response}
\label{Imaging}

\begin{figure}[ht]
  \centering
  \includegraphics[width=0.80\textwidth]{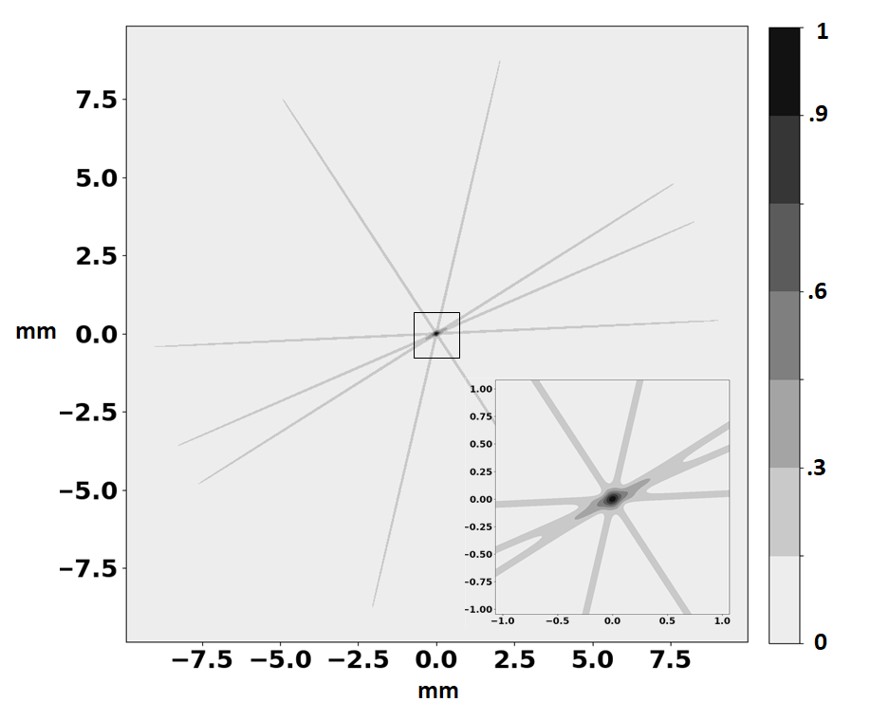}
  \caption{An illustration of image reconstruction from a small number of LORs.
Each `needle' is characterized by Gaussian distributions in the
transverse and longitudinal directions, with $\sigma_T\approx 40$
microns and $\sigma_L\approx 1.2$ cm, with a volume of $\approx 0.02$
mm$^3$. The intersection of many such volumes is dominated entirely by
$\sigma_T$, with a volume of $\approx \sigma_T^3 = 40^3$
microns$^3$.}
  \label{fig:needles}
\end{figure}

Figure~\ref{fig:needles} shows a 1-sigma surface of the probable
location of the positron-electron annihilation for a small number of
events. The transverse resolution is set by the resolution on the entry
points of each gamma ray, on the order of 40 microns when the first
cluster (efficiency $\epsilon = 85\%$) and gamma-electron scattering
location in the cluster (efficiency $\epsilon = 95\%$) are identified
from the kinematics. The overall efficiency for correct identification
of both gamma interaction points is $\approx 65\%$. The resolution
along the LOR axis is larger than the transverse resolution by a factor
of $\approx 12000/40=300$. The shape for one event is like that of a
narrow needle with Gaussian projections.


For images with many bright features close to each other, the needle
stacking produces a low-frequency background due to pile-up of needle
crossings in high density areas in the image. There are a number of
sophisticated algorithms in both image space and the corresponding
Fourier space for filtering and performing likelihood fits to PET
scanner images~\cite{Sjors_Scheres_cleaning_lecture_2017}. In the case that the detector resolution is smaller than the voxel
size, a series of ortho-normal functions can be used to subtract
low-frequency pile-up directly from the image. 

\begin{figure}[ht]
  \centering
  \includegraphics[width=0.80\textwidth]{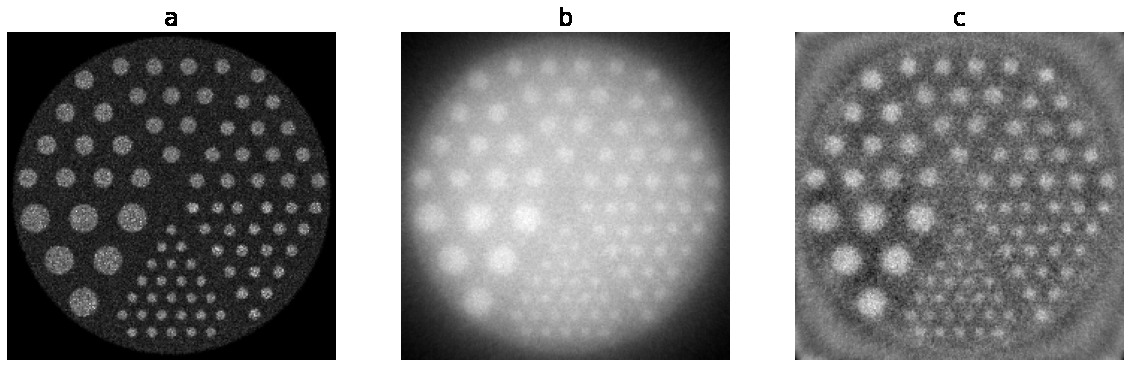}
  \caption{Imaging the Derenzo phantom~\cite{Derenzo_phantom}  with the unfiltered
  back-projection of the needle
  stack. Left: an 1 mm-thick slice through the input source distribution density of annihilations with a
  10-minute exposure using the reduced dose of 30 Bq of $^{18}$F in the rods and
  10 Bq in the surrounding volume~\cite{Ultra_Low_Dose}. The white dots represent voxels with
  annihilations.
  Middle: the reconstructed projected image from summing the weights of needles crossing each
  voxel. Right: the reconstructed projected image after subtracting a least-mean-squares fit
  to the 136 terms of an expansion in Zernike functions up to order 15.
  The detector accepts all annihilations regardless of angle and no efficiencies or in-patient absorption or scattering corrections have been applied.}
\label{fig:ZernikeFitDerenzo}
\end{figure}

The left-hand panel of Figure~\ref{fig:ZernikeFitDerenzo} shows an image of simulated annihilation locations for a 10 minute exposure of the Derenzo phantom~\cite{Derenzo_phantom} loaded with the low dose of $^{18}$F at 30 Bq/cc in the source (rod) regions and at 10 Bq/cc in the surrounding
background volume~\cite{Ultra_Low_Dose}. The detector accepts all annihilations regardless of angle, and no efficiencies or in-patient absorption or scattering corrections have been applied. To model the raw image on the left, for each of the simulated annihilations we generate a random gamma emission direction to determine a LOR, and then randomly shift the LOR according to the detector resolution (see fig. ~\ref{fig:needles}). The LORs are then added together, and a 1 mm slice is taken across the phantom to generate the image.

The image reconstructed from
the density of needles in each voxel is shown before and after
subtracting a fit to the 136 terms of the Zernike functions up to order
15~\cite{Zernike} in the middle and right-hand panels. No detector
efficiencies or in-patient absorption/scattering corrections have been
applied. More details and a table of signal-to-noise ratios are given
in Appendix A.


The TOF information from the initial scintillation light contains the
positions and times of the individual photons, with intrinsic sub-mm
resolution in the detector plane. Including the TOF information in the
determination of the initial gamma direction from the list of cluster
locations is sufficiently complex to provide an opportunity for advanced likelihood techniques such as machine learning~\cite{machine_learning_in_PET}.

\section{Summary and Conclusions}
\label{Summary}

We propose the development of PET scanners that use
ionization-activated multi-state low-Z detector media to achieve
spatial resolutions limited by the underlying physics rather than by
detector segmentation. A gamma from the positron annihilation produces
a chain of successive Compton scatterings in the detector medium,
activating the medium along the mm-scale paths of the Compton recoil
electrons. Measurement of the individual scattering locations,
deposited energies, and recoil electron directions allows using the
kinematical constraints of Compton scattering to perform a statistical
time-ordering of the scatterings. The locations of the first
interactions of both gammas determine the Line-of-Response with a
transverse rms determined by the precision of locating the start of the
corresponding recoil electron track.  A simple least-mean squares fit
to simulated events in an organic scintillator identifies the initial
scattering location of a gamma $>$85\% of the time. Mistaken
identifications typically have rms transverse resolutions
orders-of-magnitude larger, and contribute to the image as a
low-frequency low-contrast background, removable by filtering.

A time-of-flight system based on large-area MCP-PMTs provides a stable
real-time coincidence event trigger with a coincidence window on the
order of several nanoseconds. With the single-photon resolution below
40 psec of MCP-PMTs, this can be substantially reduced using the
combined data of real-time photon times and positions  and the off-line
analysis of the locations and directions of the Compton-chain
electrons, with a time-base provided by a commercial multi-channel
system with sub-10 psec resolution over the
system~\cite{Seven_Solutions}.

 As an example of a persistent low-Z medium capable of high spatial and energy resolution,
 we have simulated a two-state photoswitchable organic dye~\cite{Eric_CPAD},
 activated to a fluorescent-capable state by the
ionization of the recoil electrons. The activated state can be
optically excited multiple times to image individual activated
molecules. Energy resolution is provided by counting the activated
molecules. Location along the LOR is implemented by large-area
time-of-flight  MCP-PMT photodetectors~\cite{Incom_production}
with single photon time resolution in the tens of ps and deep sub-mm
spatial resolution.

Simulations of the Derenzo PET imaging phantom indicate a substantial
reduction in dose is possible. The 30/10 Bq dose that produces
signal-to-noise figures greater than 3 in the simulation is lower than
a typical 30/10 KBq~\cite{Ultra_Low_Dose} by a factor of 1000. However
no detector efficiencies or corrections for in-patient
absorption/scattering have been applied; these will lower the dose
reduction factor, but depend on details of the low-Z medium used and
the detector. The concept seems promising, but a quantitative estimate
of dose reduction is not yet possible.

 The concepts described here  will require development
efforts on photoswitchable organic fluorophores
(`Switchillator;)~\cite{Eric_CPAD} and other ionization-retaining
media, large-scale production of low-cost robust `commodity' large-area
psec MCP-PMTs~\cite{history_paper} and high-resolution pad-based
anodes, and economical 10 GS/sec or higher
wave-form sampling electronics systems such as the PSEC4
system~\cite{Oberla_Clermont_2014,OTPC_paper}. However achieving doses
a factor of 100 lower or more with spatial resolutions set by the
intrinsic limitations due to the underlying physics would have an
impact on public health world-wide.
%
%

\section{Acknowledgments}
We thank Andrew Hanson for advice and references on imaging. Mary
Heintz provided crucial technical and computational support. We are
grateful to Michael Grosse of the UC Physical Sciences Division for
financial support of the medical imaging aspects and students. We
especially thank Helmut Marsiske and the DOE for the long-time support
of the development of the LAPPD photodetector and associated 10 GS/sec
wave-form sampling ASIC systems for HEP experiments.

 A. Squires was supported by the Neubauer Family Foundation and the University of Chicago Materials Research Science and Engineering Center, which is funded by the National Science Foundation under award number DMR-2011854. P.
 La Riviere was partially supported from NIH R01EB026300. 
 The Geant4 simulation and code development by A. Elagin for neutrinoless double-beta decay
 were supported by the DOE under contract DE-SC0008172. Materials and supplies
 and support for drafting graphics were provided by the Physical
 Sciences Division (PSD) of the University of Chicago. The student authors were
 supported by the University's Enrico Fermi Institute, the College, and the PSD.
 This work made use of the shared facilities at the University of Chicago Materials Research Science and Engineering Center, supported by National Science Foundation under award number DMR-2011854

\section*{Appendix A: Signal-to-Noise Ratios }

The low-frequency background of the image from the 30/10 Bq/cc
simulation of the Derenzo phantom shown in
Figure~\ref{fig:ZernikeFitDerenzo} was fitted using the first 15 orders
of Zernike polynomials, comprising 136 terms. After removing the fitted
background from the image, the average signal-to-noise (S/N) ratio and
FWHM of the rods were determined as follows.

The Derenzo phantom comprises six angular slices, each containing
identical rods ranging from 6.4 mm to 19.1 mm in diameter. The rods in
each slice were fit to a model rod in a 4-parameter fit over the
1-mm$^2$ image pixels.  Each of the rod sizes was modelled as a
cylindrical trapezoid using 4 parameters: 1) the diameter of a disk of
constant value around the rod center; 2) the average background in the
angular slice excluding the rods; 3) the average value in the center
disk, and 4) the slope of a cone extending from the edge of the center
disk down to the background value.

The signal (S) in each slice was calculated as the difference between
the value of the central disk and the background.  The noise (N) was
measured as the standard deviation of the background pixels. The FWHM
was calculated using the trapezoidal shape with the height defined by
the value of the center disk, and the sloped sides extending down to
the background value. The signal-to-noise ratios S/N and FWHM values
are presented in Table~\ref{tab:SNR_and_FWHM}.

\begin{table}[hb]
\centering
\begin{tabular}{|c|c|c|c|}
\hline
Rod Diameter (mm)& Number of Rods & S/N Ratio & FWHM (mm) \\
\hline
19.1 &  6 & 4.4 & 18.0 \\
\hline
12.7 & 10 & 4.0 & 12.5 \\
\hline
11.1 & 10 & 3.6 & 10.9 \\
\hline 9.5 & 15 & 3.5 & 9.6 \\
\hline
7.5 & 21 & 3.5 & 7.2 \\
\hline
6.4 & 26 & 3.2 & 6.0 \\
\hline
 \end{tabular}
 \caption{The signal-to-noise ratios and the value of the FWHM
 determined from the fit to the image of the Derenzo phantom shown in
Figure~~\ref{fig:ZernikeFitDerenzo}. The
 six rows correspond to the six angular regions of identical rods.
The difference between the fitted FWHM and the true rod diameter is
comparable to or less than the 1 mm pixel size of the image.}
 \label{tab:SNR_and_FWHM}
\end{table}


\end{document}